\def\bq{\begin{equation}}
\def\eq{\end{equation}}
\def\bqa{\begin{eqnarray}}
\def\eqa{\end{eqnarray}}
\def\bqb{\begin{eqnarray*}}
\def\eqb{\end{eqnarray*}}

\def\simge{{\ \lower2pt\hbox{$\sim$ }\mkern-18mu \raise2pt
\hbox{$>$}\ }}

\def\simle{{\ \lower2pt\hbox{$\sim$ }\mkern-18mu \raise2pt
\hbox{$<$}\ }}

\documentclass[12pt]{article}\textwidth 16cm
\usepackage{epsfig}
\def\pr#1#2#3{ Phys. Rev. ${\bf{#1}}$ (#2) #3}

\def\pl#1#2#3{ Phys. Lett. ${\bf{#1}}$ (#2) #3 }

\def\np#1#2#3{ Nucl. Phys. ${\bf{#1}}$ (#2) #3}
\def\zp#1#2#3{ Z. Phys. ${\bf{#1}}$ (#2) #3}

\global\nulldelimiterspace = 0pt

\def\Bsl{\hbox{/\kern-.6700em$B$}} 
\def\Dsl{\hbox{/\kern-.6700em$D$}} 
\def\Wsl{\hbox{/\kern-.6700em$W$}} 

\def\roughly#1{\mathrel{\raise.3ex
    \hbox{$#1$\kern-.75em\lower1ex\hbox{$\sim$}}}}
\def\lsim{\roughly<}

\def\mh2{m^2_H}

\begin{document}
\pagenumbering{arabic}
\thispagestyle{empty}
\hspace {-0.8cm} PM/96--25 \\
\hspace {-0.8cm} FNT/T--96/18 \\
\hspace {-0.8cm} September 1996\\
\vspace {0.8cm}\\

\begin{center}
{\Large\bf Observability at LEP2 hadronic channels} \\
\vskip 12pt
{\Large\bf of a Z' with small lepton couplings} \\
 \vspace{1.8cm}
{\large  G. Montagna$^{a,b}$, F. Piccinini$^{b,a}$} \\
{\large  J. Layssac$^c$, F.M. Renard$^c$}\\
{\large  C. Verzegnassi$^d$}
\vspace {1cm}  \\

$^a$Dipartimento di Fisica Nucleare e Teorica, Universit\`a di Pavia\\
Via A.Bassi 6, 27100, Pavia, Italy.\\
\vspace{0.2cm}
$^b$ INFN, Sezione di Pavia, Via A.Bassi 6, 27100, Pavia, Italy.\\
\vspace{0.2cm}
$^c$Physique Math\'{e}matique et Th\'{e}orique, UPRES-A 5032,\\
Universit\'{e} de Montpellier II,
 F-34095 Montpellier Cedex 5.\\
\vspace{0.2cm}
$^d$ Dipartimento di Fisica,
Universit\`{a} di Lecce \\
CP193 Via Arnesano, I-73100 Lecce, \\
and INFN, Sezione di Lecce, Italy.\\
\vspace{1.5cm}

{\bf Abstract}
\end{center}
\noindent
We consider the effects of a number of models with one extra $Z$, 
with enhanced couplings to quarks, in the final 
\underline{hadronic} channels at LEP2. We show that, 
for a number of representative cases, visible effects could be 
produced even for very low values of the 
\underline{lepton} couplings, much smaller than the existing LEP1/SLC 
and the future LEP2 (lepton channel) bounds. 
\vspace{1cm}

\setcounter{page}{0}
\def\thefootnote{\arabic{footnote}}
\setcounter{footnote}{0}
\clearpage

\section{Introduction}

It has been recently suggested~\cite{one},\cite{two} that a model 
with one extra $Z \equiv Z'$ with enhanced, 
family independent couplings to quarks, 
would explain in a simple way possible departures 
from the SM predictions for the $Z$ couplings to heavy quarks
~\cite{three} 
and, at the same time, a possible excess
of dijet events  at CDF~\cite{four}. For non vanishing values 
of its charged lepton couplings, it would be possible to 
see the effects of such a $Z'$ (``hadrophilic'' in the notation 
of Ref.~\cite{one}) in the final hadronic channels at LEP2~\cite{one}
 under the expected experimental accuracy for the relevant 
observables~\cite{five}. 

Following the original proposal of Refs.~\cite{one},
\cite{two}, several papers have appeared~\cite{six},\cite{seven}
 that investigate the possibility of finding a realistic 
theoretical model
 where this special $Z'$ can be accomodated without introducing 
 unwanted anomalies. In the original proposal of Ref.~\cite{six}, 
this goal was achieved assuming that the couplings of the $Z'$
 with leptons were (``approximately'') vanishing. This generated 
the word ``leptophobic'' $Z'$, the ``leptophobia''
 syndrome being substantially supported by the phenomenological 
evidence that leptonic widths and asymmetries are in very good 
agreement with the SM predictions and do not necessitate, in 
principle, any correction of New Physics type. \par
 A rather fundamental detail that we want to stress at this 
point is that, 
from an experimental point of view, there is a dramatic 
 difference between a $Z'$ with ``approximately'' 
vanishing lepton couplings and one with \underline{exactly}
 vanishing lepton couplings. The latter one, strictly speaking, 
 could not be revealed by the conventional techniques, based 
on searches of virtual $Z'$ effects, at future $e^+ e^-$ 
colliders~\cite{eight},\cite{nine}, so that LEP2 and NLC 
would be ``blind'' to such an effect, even for very low 
values of the $Z'$ mass. On the contrary, 
a $Z'$ with ``small'' lepton couplings might 
compensate this kind of weakness with a suitable enhancement 
of its quark couplings. In this case, the final 
\underline{hadronic} channels at LEP2 and NLC might be able 
to evidentiate the hadrophilic $Z'$ virtual 
effect in the conventional way, i.e. via small deviations 
in the hadronic cross section and/or asymmetries, even 
 for substantial (in the LEP2 case, of about 1~TeV) values 
of the $Z'$ mass. \par

Since the point that we have just mentioned seems to us worth of 
being investigated in more detail, we have decided to devote this 
paper to a discussion of the ``minimal'' size of the lepton 
 couplings that would still make a hadrophilic $Z'$ visible at LEP2 
\footnote{and, therefore, not ``LEP2--phobic'', but rather, 
in a generalized 
notation, ``LEP2--philic''}. \par

The outline of the paper is as follows. We describe in Section 2 a 
 theoretical model where one $Z'$ exists, with particular emphasis 
on its phenomenological inputs and constraints. In section 3 
we present the 
implementation of the QED radiative corrections in the 
calculation of the $Z'$ effects on the hadronic observables. The discussion 
of the numerical results is given in Section 4 and the main conclusions
 are summarized in Section 5. 

\section{Phenomenological constraints and theoretical model}

Since we shall assume that $Z'$ quark couplings are ``large'', we shall 
introduce and use from now on four parameters that are indicative of this 
relative enhancement, more precisely the ratios 

\bq \xi_{Vf} \equiv {g'_{Vf}\over g_{Vf}} 
\qquad ; \qquad
\xi_{Af} \equiv {g'_{Af}\over g_{Af}}   
\label{eq:csiq}
\eq

\noindent
where $g_{V,Af} (f= u,d)$ are the usual $Z f \bar f $ couplings, 
$g_{Vf}= I_{3L}^f - 2 s^2_{eff} Q^f$, where $s^2_{eff}=0.232$ and 
$g_{Af}= I_{3L}^f$. The $Z'$ effect 
in the hadronic channels at LEP2 will depend on six effective parameters, 
those of eq.~(\ref{eq:csiq}) and 
the conventionally defined $Z'$ lepton couplings 
$g'_{Vl}$ and $g'_{Al}$ (we follow the convention of Ref.~\cite{one}), each one 
divided by $\sqrt{M^2_{Z'} - q^2}$, where $q^2$ is the total c.m. squared 
energy. We shall assume at this stage that the $Z'$ couplings are 
family independent. 

To give a hint of the practical situation, we now 
write the approximate expressions (computed at the lowest order ``effective'' 
Born level) of the relative shifts from the SM predictions 
(due to a general $Z'$) of the three experimental hadronic quantities 
that will be realistically measured at LEP2, i.e. the total hadronic 
cross section $\sigma_5$ and the $b \bar b$ cross section and 
forward--backward asymmetry $\sigma_b$, $A_{FB,b}$. For these observables 
it is not difficult to derive the following expressions for the 
aforementioned shifts, that read, in the chosen configuration 
$\sqrt{q^2} = 175$~GeV (where we shall work from now on)~\cite{ten} and for 
large $M_{Z'}$ values, $M_{Z'}^2 \gg M_Z^2$ (chosen in this illustrative 
example for pure simplicity reasons):

\bqa  
{ {\delta \sigma_5^{(Z')} } \over  {\sigma_5} }  \, \simeq
\, \left[1.4 \times 10^{-2} \right] \left( { {1 TeV} \over {M_Z'} }\right)^2 
\! \! \! & \! \! \! \Big\{g'_{Vl} \left[0.67 \, \xi_{Vu} + 0.83 \, \xi_{Vd} + 
0.232 \, \xi_{Au} + 0.41 \, \xi_{Ad} \right] \nonumber \\
&+ g'_{Al} \left[1.64 \, \xi_{Ad} + 1,10 \, \xi_{Au} + 0.17 \, \xi_{Vu} 
+ 0.84 \, \xi_{Vd} \right] \Big\}  
\label{eq:sig5}
\eqa

\bq  
{ {\delta \sigma_b^{(Z')} } \over  {\sigma_b} }  \, \simeq
\, \left[5 \times 10^{-2} \right] \left( { {1 TeV} \over {M_Z'} }\right)^2 
\Big\{g'_{Vl} \left[0.50 \, \xi_{Vd} + 0.19 \, \xi_{Ad} \right] 
+ g'_{Al} \left[0.96 \, \xi_{Ad} + 0.46 \, \xi_{Vd} \right] \Big\}  \\
\label{eq:sigb}
\eq

\bq  
{ {\delta A_{FB,b}^{(Z')} } \over  {A_{FB,b}} }  \, \simeq
\, \left[10 \times 10^{-2} \right] \left( { {1 TeV} \over {M_Z'} }\right)^2 
\Big\{g'_{Vl} \left[0.21 \, \xi_{Vd} + 0.40 \, \xi_{Ad} \right] 
+ g'_{Al} \left[-0.23 \, \xi_{Vd} - 0.05 \, \xi_{Ad} \right] \Big\}  \\
\label{eq:afbb}
\eq

Eqs.~(\ref{eq:sig5})--(\ref{eq:afbb})
can be easily derived within a special theoretical framework, 
denoted ``$Z$--peak subtracted'' 
representation of four--fermion processes~\cite{eleven},\cite{twelve}. 
In fact, the complete relevant expression for general $q^2$ and $M_{Z'}$ values 
can be found in Ref.~\cite{twelve}, and we shall not insist here on their 
derivation, particularly since a more complete evaluation of the effects, 
that takes the dangerous QED radiation properly into account, will be 
performed in this paper. The reason why we showed these approximate expressions 
is that from their inspection a certain number of general conclusions can be 
already drawn. In particular: 

I) the numerical square brackets that appear as first terms 
of the r.h.s. of eqs.~(\ref{eq:sig5})--(\ref{eq:afbb}) 
are the expected experimental accuracies for the 
various quantities (two standard deviations) at LEP2. 
In order to make a 
visible effect, all the quantities in the three curly brackets must be 
(at least) of order one. For values of $| \xi_i |$ reasonably larger
than one (say, of about $\sim 3$ to 
$\sim 4$) one sees that e.g. values of 
$g'_{Vl} = g'_{Al} \simeq \pm \, 0.1$ ($M_{Z'} / 1$~TeV) would produce 
a visible effect in $\sigma_5$ and also in $\sigma_b$. These indicative 
results will be substantially confirmed by the more 
accurate determination 
that we shall perform. \par

II) The three experimental observables of 
eqs.~(\ref{eq:sig5})--(\ref{eq:afbb}) 
clearly exhibit a different 
dependence on the four quark ratios $\xi_i$. This 
means that a certain non 
trivial correlation will exist between the three 
``$Z'$ shifts'' that might 
be useful to differentiate a model with a $Z'$ of 
this type from other possible 
competitor models of New Physics by looking at 
its possible combined effects in 
the hadronic channels at LEP2. \par

III) In order to perform a reasonably simple search of the 
aforementioned effects, 
a certain reduction of the number of $Z'$ parameters must 
be enforced. This 
can be achieved either by imposing a number of purely phenomenological 
restrictions or by selecting a specific model. We shall 
immediately illustrate 
possible ways of carrying on this program in what follows. 

A first simple way to eliminate one of the four quark 
parameters $\xi_i$ is that of using the available 
experimental information on the $Z$ hadronic width $\Gamma_h$. In our 
analysis, any possible deviation of a given observable from the 
corresponding SM prediction is interpreted as a ``$Z'$ shift''. 
For $\Gamma_h$
 the most recent analyses~\cite{three} give:

\bq { {\Gamma_h^{(exp)} - \Gamma_h^{(SM)}} 
\over {\Gamma_h^{(SM)}} } \, = \, +0.0005 \, \pm \, 0.0014 . 
\label{eq:gammah}
\eq

In the situation that we are considering, 
the l.h.s. of eq.~(\ref{eq:gammah}) can also be written as 
$\delta \Gamma_h^{(Z')} / \Gamma_h$. For this quantity one can 
write, neglecting for simplicity terms of higher order in 
``small'' parameters (like the $Z-Z'$ mixing angle $\vartheta_M$, 
defined in the conventional way):

\bq { {\delta \Gamma_h^{(Z')}} \over {\Gamma_h} } 
\, \simeq \, {3 \over 2} 
\, \delta_{\rho}^{(Z')} + {36 \over 59} \, \vartheta_M 
\left[ \xi_{Au} + 4 \, g_{vu}^2 \, \xi_{Vu} + 
{3 \over 2} \, \xi_{Ad} + 6 \, g_{vd}^2 \, \xi_{Vd} \right] 
\label{eq:dgammah}
\eq
\noindent
where, for reasonably large $Z'$ masses:

\bq \delta_\rho^{(Z')} \simeq \vartheta_M^2 
\, {{M_{Z'}^2} \over{M_Z^2} } \, . 
\label{eq:link}
\eq

Imposing the equality of eqs.~(\ref{eq:gammah}) 
and (\ref{eq:dgammah}) gives rise to a linear 
relationship between the four parameters $\xi_i$, that reads:

\bq \xi_{Au} = - \, {3 \over 2} \, \xi_{Ad} -6 \, g_{vd}^2 \, \xi_{Vd} 
-4 \, g_{vu}^2 \, \xi_{Vu} \pm \delta 
\label{eq:relation}
\eq
\noindent
where $\delta$ depends on $M_{Z'}$, $\delta_{\rho}^{(Z')}$ 
and the experimental result in the r.h.s. of eq.~(\ref{eq:gammah}). 
Assuming for $\delta_{\rho}^{(Z')}$ the reasonable 
bound~\cite{fourtheen}:

\bq \delta_{\rho}^{(Z')} \simle 3 \times 10^{-3} 
\label{eq:dbound}
\eq
\noindent
one sees that the contribution to $\delta$ of 
the experimental input $\sim \left(\Gamma_h^{(exp)}-\Gamma_h^{(SM)} \right)$ 
can be essentially ignored in the region of large 
$|\xi_{Au}|$ in which we are interested, so that one can write for
practical purposes:

\bq \delta \, \simeq \, {5 \over 2}  {{M_{Z'}} \over {M_Z} } 
\sqrt{\delta_{\rho}^{(Z')}} \, \simeq \, 27 \left( 
{M_{Z'} \over {1 TeV} } \right) \sqrt{\delta_{\rho}^{(Z')} } .  
\label{eq:delta}
\eq

The previous quantity is such that, in the region of interest for LEP2, 
 $M_{Z'} \leq 1$TeV, and for the worst assumption on 
$\delta_{\rho}^{(Z')}$, 
\bq |\delta| \simle 1 \, . \eq 

A shift like that of eq.~(\ref{eq:delta}) would therefore, at most, change the 
 value of $\xi_{Au}$ (for fixed $\xi_{Vu}, \xi_{Ad}, \xi_{Vd}$) 
by an amount sensibly smaller than the ``relevant'' 
($|\xi_{Au}| \simge 3$) ones. In practice, and for simplicity reasons, 
we shall incorporate from now on the parameter $\delta$ of 
eq.~(\ref{eq:relation}) into a generalized definition of $\xi_{Au}$, that 
will read:

\bq \xi^{'}_{Au} \equiv \xi_{Au} \pm \delta 
\equiv -{3 \over 2} \, \xi_{Ad} -6 \, g_{vd}^2 \, \xi_{Vd} 
-4 \, g_{vu}^2 \, \xi_{Vu} 
\label{eq:csief}
\eq
\noindent
and would be practically identical with the conventional $\xi_{Au}$ 
in the ``large'' $\xi_{Au}$ 
region for $M_{Z'}$ sufficiently below one TeV or 
$\delta_{\rho}^{(Z')}$ sufficiently smaller 
than the bound of eq.(9).

As a second constraint to be used in our analysis, we shall use the 
 condition that can be derived from the request (that we shall maintain 
in this paper) that the considered $Z'$ is ``reasonably'' narrow. In 
practice, we shall make this statement more quantitative by demanding that
\bq { {\Gamma_{Z'} } \over {M_{Z'}} } \simle 0.20 . 
\label{eq:ubound}
\eq

From the expression of the $Z'$ width given e.g. in Ref.~\cite{one} 
one sees that eq.~(\ref{eq:ubound}) leads to  the constraint

\bq { {3 \alpha} \over {4 s^2_{eff} c^2_{eff}}} 
\left[{1 \over 4} \, (\xi_{Au}^2 + \xi_{Ad}^2) 
+ g_{Vu}^2 \, \xi_{Vu}^2 + g_{Vd}^2 \, \xi_{Vd}^2 \right] \simle 0.20 
\eq

\noindent
that becomes numerically ($s^2_{eff} = 1 - c^2_{eff} = 0.232$)

\bq  \left[ \xi_{Au}^2 + \xi_{Ad}^2 + 0.48 \,
\xi_{Vd}^2 + 0.15 \, \xi_{Vu}^2 \right] \simle 26 .  
\label{eq:eubound}
\eq

Eqs.~(\ref{eq:relation}) and (\ref{eq:eubound})
are the two purely phenomenological constraints 
that we shall use in our analysis. To further reduce 
the number of parameters, extra theoretical assumptions 
must be enforced. One natural possibility, that has been suggested 
in Ref.~\cite{two}, is that of assuming that the $Z'$ is associated 
to an extra $U(1)$ that commutes with the standard 
$SU(2)_L \bigotimes U(1)_Y$ group. In terms of the $Z'$ quark
effective couplings, this means that $g'_{uL} = g'_{dL}$ and therefore
 that 
\bq g_{vu} \, \xi_{Vu} + {1 \over 2} \, \xi_{Au} 
= g_{vd} \, \xi_{Vd} - {1 \over 2} \, \xi_{Ad} \, . 
\label{eq:singlet}
\eq

Following the suggestion of Ref.~\cite{two}, and assuming that anomalies 
can be cancelled by a proper ad hoc mechanism that is beyond the 
purposes of this analysis, we shall now investigate the visible properties 
at LEP2 of a model that meets the constraints eqs.~(\ref{eq:relation}),
(\ref{eq:eubound}) and
(\ref{eq:singlet}) and that we shall call from now on model $A$. 
Briefly, we shall perform the following operations: \par\noindent
I) we shall concentrate on the experimental variables 
$\sigma_b = \sigma(e^+ e^- \to b \bar b), 
A_{FB,b}$ (the forward-backward asymmetry for 
$b \bar b$ production) and $\sigma_5$ (the cross section for 
hadron production at LEP2). We shall work assuming a total 
c.m. energy of 175~GeV and a total integrated luminosity 
of $500$~pb$^{-1}$, following the discussion on the expected 
experimental accuracies given in Ref.~\cite{ten}. In particular, the 
\underline{relative} experimental errors on $\sigma_5,\sigma_b$ and 
$A_{FB,b}$ will be taken as $0.7\%, 2.5\%$ and $5\%$, respectively. 
\par\noindent
II) The $Z'$ parameters will be chosen in the following way: 
the two $b$ effective couplings $\xi_{Vb}$, $\xi_{Ab}$ 
will be left free to vary, within the allowed region limited 
by the bound imposed by the request of small $Z'$ width, 
eq.~(\ref{eq:eubound}).
The two  $u$ couplings 
$\xi_{Au}, \xi_{Vu}$ will be fixed 
by the conditions of eqs.~(\ref{eq:relation}) and (\ref{eq:singlet}). 
In practice this sets a limitation on $|\xi_i|$ of the kind 
$|\xi_i| \lsim$ 3---4. We shall neglect for 
a first investigation the extra 
parameter $\delta$, remembering from our previous 
discussion that values of 
$\delta$ below the limit $|\delta| \simle 1$ will 
not change the relevant 
conclusions (as we checked for intermediate $\delta$ 
values) but will be 
simply reabsorbed into a redefinition of an effective $\xi_{Au}'$ 
parameter, eq.~(\ref{eq:csief}). 
For purely orientative reasons, $M_{Z'}$ has been 
fixed at a value $M_{Z'} = 800$~GeV that was 
suggested by our analysis of
the CDF events~\cite{one}; rescaling the 
results to different $M_{Z'}$ 
values is trivial as suggested by 
eqs.~(\ref{eq:sig5})--(\ref{eq:afbb}). 
\par\noindent
III) The two $Z'$ lepton parameters $g_{Vl}', g_{Al}'$ have been 
analyzed in the following way. We have set
\bq r = { {g_{Al}'} \over  {g_{Vl}'} } 
\label{eq:rdef}
\eq

\noindent
and considered different cases at fixed $r$. In particular, 
we have analysed the special situations $|r| \gg 1$ (predominant 
axial $Z'$ lepton coupling), $r=0$ (predominant
vector coupling), $r = \pm 1$ (left and right-handed couplings); these 
cases should give a picture of various possibilities that might 
reasonably occur in a class of theoretical models. 
\par\noindent
IV) Once the values of $r$ and of $M_{Z'}$ are fixed, the three  
experimental quantities that we have considered only depend on 
\underline{two} residual effective parameters 
$\sim g'_{Vl} \cdot \xi_{Vb}$, $g'_{Vl} \cdot \xi_{Ab}$. 
This means that the three relative $Z'$ shifts on the observables will 
be related by a certain relationship that will not depend on the 
previous residual parameters, and will therefore lie on a certain $3$d 
region fully characteristic of the $Z'$ model. Following the notation 
of a previous work~\cite{fifteen}, we have called ``$Z'$ reservation'' this 
region and have computed it for the various $r$ values. The aim of this 
operation was also that of showing that certain possible 
competitor theoretical models will generate different ``reservations'', which 
will be shown at the end of this paper. 

\section{Inclusion of QED corrections}

The practical evaluation of the $Z'$ shift must necessarily take 
into account the potentially dangerous QED radiation effects. In order 
to accomplish this goal, we have 
upgraded a Monte Carlo program, previously used for an 
 analysis of the $Z'$ effects on the leptonic observables 
at LEP2~\cite{fifteen}, 
 to give predictions for the relevant hadronic 
observables $\sigma_b$, $A_{FB,b}$ 
and $\sigma_5$, keeping under control the bulk of the contribution due 
to the emission of soft and collinear 
(undetected) initial--state photonic radiation.
 In order to assess the normalization
of the $Z'$ shifts as returned by the 
program, the predictions for the hadronic SM quantities 
 have been compared with the high--precision theoretical results 
of typically used LEP software~\cite{sixteen}
 and found to be in agreement at the level of few per cent
for the integrated cross sections and within 1\% for the $b$-asymmetry.
However, these small discrepancies, which have to be essentially ascribed 
to neglecting final-state QCD corrections, mixed 
electroweak--QCD contributions and finite-mass 
 effects in our calculation, marginally affect the relevant conclusions 
 because the neglected short--distance corrections largely cancel out 
in the numerical evaluation of the \underline{relative} shifts in which 
we are interested. 

Following the strategy already employed in Ref.~\cite{fifteen}, 
the $Z'$ contribution has been included in the 
lowest--order 
calculation of the cross section implemented in the Monte Carlo computing the
$s-$channel amplitudes associated
 to the production of a $q {\bar q}$ (massless) pair in
$e^+ e^-$ annihilations mediated by the exchange of a photon, a
standard $Z$ and an additional $Z'$ boson.
The resulting cross section has been dressed with 
leading weak corrections included in the form of Improved Born 
Approximation and the potentially dangerous distortions introduced by 
initial--state QED radiation have been kept under
control employing the QED structure function approach~\cite{seventeen},
i.e. convoluting the short--distance cross section 
$\sigma_0$ with electron(positron) structure functions as follows
~\footnote{\footnotesize The actual implementation of
QED corrections is performed, in the Monte Carlo code, at the level of the
differential cross section, taking into account all the relevant kinematical
effects according e.g. 
to Ref.~\cite{eighteen}; in the present paper only a simplified 
formula is described, for the sake of simplicity. }
\begin{eqnarray}
\sigma (q^2) = \int d x_1 \, d x_2
\, D(x_1,q^2) D(x_2,q^2) \sigma_0 \left( x_1 x_2 q^2 \right) 
\Theta({\rm cuts}),
\label{eq:master}
\end{eqnarray}
where $D(x,q^2)$ is the electron (positron) structure function (whose
 typical expression can be found elsewhere~\cite{seventeen,eighteen}) 
and 
 $\Theta({\rm cuts})$ stands for the rejection algorithm
to implement possible experimental cuts. In our analysis, 
 we use the cut $s' / q^2 > 0.35$ (where $s' =
x_1 x_2 q^2$ is the invariant mass of the event 
after initial-state radiation)  
 imposed in order to avoid, 
according to standard LEP2 selection criteria, 
the unwanted events due to $Z$ radiative return 
and hence disentangle the interesting virtual $Z^{'}$ effects.

The numerical results obtained via the Monte Carlo program 
have been compared with the analytical formulae given in
eqs.~(\ref{eq:sig5})--(\ref{eq:afbb}), 
in order to estimate the size of those Born level contributions 
 neglected in the derivation of the
analytical expressions for the $Z'$ shifts 
written in those equations and taken into account 
in the numerical simulation.
For simplicity reasons, this comparison has been performed 
assuming the relations among 
the couplings valid for model $A$ 
but neglecting 
the QED corrections. The content of this analysis 
is summarized in Fig.~1 for the illustrative case $r = -1$. 
In this plot one can see:
the results for the three hadronic observables as predicted by the 
analytical formulae of eqs.~(\ref{eq:sig5})--(\ref{eq:afbb}) (Fig.~1a); 
the numerical results obtained with the Monte Carlo code, switching off 
both the contribution of the $Z'-$exchange squared amplitude and the 
$Z'$ width, 
the latter calculated according to the 
expression given in Ref.~\cite{one} (Fig.~1b); 
the numerical results obtained with the Monte Carlo, 
switching off the contribution of the $Z'$ width but retaining the 
$Z'$ squared amplitude (Fig.~1c); the full numerical results of the 
Monte Carlo, with both the $Z'$ squared amplitude and $Z'$ width 
switched on (Fig.~1d). As can be seen from the comparison 
between Fig.~1a and Fig.~1b, there is a rather satisfactory agreement between 
analytical and Monte Carlo numerical estimates when 
the contributions due to non--linear
effects in the $Z'$ (lepton and quark) couplings
introduced in the calculation of the relative deviations
by the $Z'$ squared amplitude and the
$Z'$ width are discarded. On the other hand, this step by step 
 investigation also points out that, in view of a
more precise evaluation of the global $Z'$ virtual effects, 
the non--linear contributions in the $Z'$ couplings 
have to be accounted for, whereas 
they can be neglected if one is interested in the order of magnitude
of the relevant shifts. These general conclusions remain valid for all
 the values of $r$ that we have considered.

\section{Numerical results and discussion}

After this preliminary check of our computational algorithm, we have drawn, 
for all the representative values of $r$, the characteristic surfaces of 
model $A$ in the 3d-space of the shifts in the three considered 
hadronic channel observables. Contrary to the previous preliminary 
investigation, we have now fully imposed both phenomenological conditions 
eq.~(\ref{eq:csief}), (\ref{eq:eubound}) and
eq.~(\ref{eq:singlet}) and the effect due to QED radiation. Moreover, 
we have been limited in our analysis to values 
of the leptonic couplings that are essentially ``small''. In particular, 
we have assumed that these couplings would not produce any observable effect 
at LEP2 in the \underline{leptonic} channels. From the previous analysis 
performed in Ref.~\cite{eight} this means that, to 95\% CL, the two $Z'$ 
couplings $g'_{Vl}, g'_{Al}$ will have to lie within a certain region (that 
depends on the fixed $M_Z'$ value) that corresponds to an ellipse in their 
plane. We shall refer to this region as to the ``leptonic channel 
ellipse'' at LEP2.

The results of our calculations for different $r$ are shown in 
Figs.~2--6. The central box in the Figures corresponds to a ``black'' region 
of non visibility, drawn for simplicity in a rather conservative way 
starting from the assumed experimental accuracies (a more 
accurate estimate of the ``black'' region would be straightforward, 
but would not change appreciably our discussion). 
Should the experimental 
point lie on one of the various reservations, 
outside the non observability 
region, it would be a reasonable indication of the existence of 
a $Z'$ of the considered type. 

Two more things should be now stressed. The first one is that the 
observability domains in the various cases are evidently 
not empty, as one 
sees from inspection. This means that for all the relevant 
values of $r$, 
and for $Z'$ leptonic couplings not visible in the LEP2 
leptonic channels, 
 there are values 
of the $Z'$ quark couplings in model A, 
consistent with the available experimental 
information 
eq.~(\ref{eq:gammah}), that would make a narrow $Z'$ 
visible in the LEP2 
hadronic channels. The second one is that our 3d--analysis 
might be able in the considered example 
to differentiate genuine $Z'$ effects from effects 
due to competitor theoretical models, owing to the particularly simple 
parameterization that is valid for the representative model $A$.

To make this statement more quantitative, we have considered the case 
 of a model where anomalous triple gauge couplings~\cite{nineteen} are 
present. We have followed the notations and the assumptions of Hagiwara 
et al.~\cite{twenty}, and worked in the framework 
of Refs.~\cite{eleven}, 
\cite{twelve}, where it has been shown that the number 
of relevant parameters
 of the model in the final two fermion channel at LEP2 is reduced 
to two~\cite{twentyone}. For this model, that we shall 
call model $B$ from 
now on, it will be therefore possible to draw, in the same 3d space of 
Figs.~2--6, an analogous surface that will be its 
characteristic feature. 
We have computed this curve only in Born approximation, 
where it can be given 
a simple analytic expression (all details can be easily derived from a 
generalization of the analysis performed in Ref.~\cite{twentyone}). 
A more accurate determination would be possible, but we know from our 
previous discussion that at least the dominant shape of the surface 
would not change. The resulting domain is shown 
in Fig.~7, and one can see 
from inspection that it is different from the domains
allowed for model A with the considered $r$ values. 
There would be thus a clean 
differentiation of model A,
from at least one reasonable 
alternative competitor model, 
provided by a careful analysis of the LEP2 hadronic channels. 

The final point that we want to discuss now is the one that was raised 
 in the introduction of our paper. Using the same 
computational algorithm 
that led us to the previous conclusions, we have tried to determine, 
for a convenient 
choice of the enhanced $Z'$ quark couplings,
 the \underline{minimum} $Z'$ lepton couplings 
that would produce a visible effect in the hadronic channels of LEP2.
 For practical reasons we have now focused our attention on the two 
 hadronic cross sections $\sigma_5$ and $\sigma_b$, 
and imposed the visibility 
condition at two sigmas for each observable, 
given the assumed experimental
precision.

In principle, our analysis could be performed allowing the $Z'$ quark 
couplings to vary in a proper domain where the three conditions 
eqs.~(\ref{eq:csief}), (\ref{eq:ubound}) and 
(\ref{eq:singlet}) are satisfied. 
This would then define a real set of minimal values for 
$g'_{Vl}, g'_{Al}$. But for the first purposes of indication 
we thought that it was enough to consider a couple of typical sets of 
$Z'$ quark couplings that met all the imposed requests and 
represented, so to say, two completely reasonable examples. In this 
spirit, we have chosen the two representative couples 
$\xi_{Vb} = \xi_{Ab} = 2$ (couple I) and $\xi_{Vb} = \xi_{Ab} = -2$
 (couple II); the remaining $Z'uu$ couplings have been fixed by the 
conditions eqs.~(\ref{eq:csief}) and (\ref{eq:singlet}), 
neglecting for simplicity the small parameter $\delta$. With 
these choices, the constraint eq.~(\ref{eq:ubound}) is met and all 
the values of the $Z'$ quark couplings lie in a reasonable 
 range, that still corresponds to a typical electroweak 
strength, as discussed in Ref.~\cite{one}. 

The results of our investigation are shown 
in Fig.~8, in correspondence to 
the special choice $M_Z' = 800$~GeV. The main 
features, in our opinion, are 
the following ones:
\par\noindent
I) the role of $\sigma_b$ and $\sigma_5$ is, in 
this search, complementary:
 the first observable is more reactive to small $Z'$ axial lepton 
 couplings, the second one to small vector ones. From the combination 
of the two measurements, a region of ($g'_{Vl}, g'_{Al}$) values
 of modulus typically around $\sim$ 0.1--0.2 appears to be still 
able to produce visible effects;
\par\noindent
II) typical values of the $Z'$ lepton couplings that would make the
$Z'$ visible 
in the final hadronic channels are much smaller 
than typical values that 
would \underline{not} make it visible in the final lepton channels (inside the 
lepton ellipse depicted in Fig.~8).

These main indications that can be derived from an inspection of 
Fig.~8 can be generalized to a more specific search, where a proper 
minimization program has been used. The results of this calculation, 
that has been performed in Born approximation for the reasons that we 
have already discussed, indicate that the minimum size of the $Z'$ 
lepton couplings in model $A$ that would still be seen in the 
combined analysis of $\sigma_b$ and $\sigma_5$ at LEP2 
(for values of the $Z'$ quark couplings satisfying our 
requests) are, for a general $Z'$ mass:
\bq 
|g'_{Vl}| \simge 0.1 \left[ { {M_{Z'}} \over {1 TeV} } \right] 
\label{eq:fr}
\eq
\bq |g'_{Al}| \simge 0.1 \left[ { {M_{Z'}} \over {1 TeV} } \right] 
\label{eq:sr}
\eq
that are approximately five times smaller than the limits 
from the lepton ellipse derived from the lepton channels at LEP2.\par
Our results eqs.~(\ref{eq:fr}), (\ref{eq:sr}) should now be 
compared with the available experimental information from LEP1/SLC. 
This can be done if an analysis of the latter information is performed 
like in Ref.~\cite{one}, i.e. allowing the $Z'$ couplings to be 
completely free (but still family independent as in our assumption). 
This leaves six parameters in charged lepton--quark sector, to be 
analyzed together with the extra parameters $\vartheta_M$ (the usual 
mixing angle) and $\delta \rho^{(Z')}$, the 
modification of the $\rho$ parameter due to a 
$Z'$, whose relationship with $\vartheta_M$ 
is given in eq.~(\ref{eq:link}). Assuming a bound for $\delta \rho^{(Z')}$
 like that of eq.~(\ref{eq:dbound}), i.e. $\sim 3 \times 10^{-3}$, 
one obtains therefore an upper bound for the mixing angle:
\bq
|\vartheta_M| \simle \sqrt{0.003} \, \, {{M_Z} \over {M_{Z'}}} .
\label{eq:thbound}
\eq

From the available LEP1/SLC data one can derive, in general, 
bounds for the product of $\vartheta_M$ with $Z'$ couplings. Assuming 
that $\vartheta_M$ saturates its bound, eq.~(\ref{eq:thbound})
 will consequently provide the \underline{minimal} bounds 
of lepton couplings. From the last communicated data
\cite{three} where both LEP1 and SLC data were used, one 
would derive at one standard deviation and for the bound on
$\delta \rho^{(Z')}$ of eq.~(\ref{eq:dbound}):
\bq
|g'_{Vl}| \simle 0.4 \left( { {M_{Z'}} \over {1 TeV} } \right) \quad 
; \quad |g'_{Al}| \simle 0.5 \left( { {M_{Z'}} \over {1 TeV} } \right) .
\eq

Note that, if a value of $\delta \rho^{(Z')}$ as low as one permille 
were assumed, the corresponding minimal values would increase 
by almost a 
factor of two.

When one repeats, under the same previous assumptions, this analysis
for the quark couplings, the corresponding minimal bands are 
in fact much 
larger, and somehow depending on which set of data (LEP1 or SLC) 
is used. Using, whenever possible, the average LEP/SLC data and 
following the same procedure as in Appendix A of Ref.\cite{one}, we have
found the following results:

\bq \eta_M \xi_{Vb} \simeq (-9.99 \pm 14.87)({M_{Z'}\over 1~ TeV}) \ \ \
 \eq

\bq \eta_M \xi_{Ab} \simeq (+5.86 \pm 7.17)({M_{Z'}\over 1~ TeV}) \ \ \
 \eq

\bq \eta_M \xi_{Vc} \simeq (+0.028 \pm 22.77)({M_{Z'}\over 1~ TeV}) \ \ \
 \eq

\bq \eta_M \xi_{Ac} \simeq (-0.74 \pm 7.28)({M_{Z'}\over 1~ TeV}) \ \ \
 \eq
\noindent
where $\eta_M=\pm 1$. As one sees, for the four effective ratios
$\xi_i$, values of order $\sim 10 [M_{Z'}/ 1~TeV]$ are still
perfectly consistent with the available experimental information.
Enforcing the further constraint of reproducing a possible excess of
dijet events at CDF\cite{four} would select values of $|\xi_i|$ of
about 3---4 as shown in Ref.\cite{one}. One sees therefore that the
values of $\xi_i$ that we used in the present paper are perfectly
consistent with all available data.

\section{Conclusions}

In conclusion, we have analyzed a rather general theoretical model where 
one $Z'$ exists, related to an extra $U(1)$ that commutes with the 
Standard electroweak group, as suggested in Ref.~\cite{two}. Imposing on the 
$Z'$ quark couplings to satisfy the single phenomenological condition 
dictated by the available experimental value of the $Z$ hadronic 
width measured at LEP1, and assuming that the $Z'$ width is reasonably 
narrow, we have taken into account a range of $Z'$ quark couplings 
that are larger than the corresponding ones for the SM $Z$, but 
perfectly consistent with all the available experimental information.

In such a picture, we have verified that the minimum values of the 
$Z'$ lepton couplings that would still give visible effects in the two 
hadronic cross sections ($\sigma_b$ and $\sigma_5$) 
measurable at LEP2 are those 
of eqs.~(\ref{eq:fr}) and (\ref{eq:sr}),
at least five times smaller than the existing bounds 
from LEP1/SLC (and also five times smaller than
the future bounds from negative searches in the final 
lepton channels at 
LEP2). For values of the $Z'$ lepton couplings $g'_{V,Al}$ larger than 
those of eqs.~(\ref{eq:fr}), (\ref{eq:sr}) we have shown 
that the effects of 
the considered model at LEP2 would be seen in the final 
hadronic channels 
in a ``reasonably'' unambiguous way, for all possible choices of the 
ratio $g'_{Vl} / g'_{Al}$. We believe thus to have somehow quantified 
the typical values of $|g'_{Vl}|,|g'_{Al}|$ below which it is allowed 
to identify a ``small'' $Z'$ lepton coupling 
with a ``vanishing'' one. For 
values of the $|g'_l|$ moduli beyond the 
$\sim 0.1 \left[M_Z' / 1 TeV \right]$
threshold, the previous identification might lead to very negative 
consequences.

\vskip 12pt\noindent
\leftline{\Large \bf Acknowledgements}
\vskip 12pt\noindent
We
would like to thank O.~Nicrosini and A.~Rotondi for useful 
discussions.


\newpage

\centerline{\Large \bf Figure Captions }

\vspace{0.5cm}\noindent
{\bf Fig.~1} {Comparison between analytical formulae and Monte Carlo 
results for the $Z'$ shifts on the hadronic 
observables. 
Analytical results according to eqs.~(2)--(4) 
(Fig.~1a); Monte Carlo results without the contributions of 
the $Z'$ squared amplitude and $Z'$ width (Fig.~1b); Monte Carlo results 
without the effect of the $Z'$ width but with the 
contribution of the $Z'$ squared 
amplitude (Fig.~1c); full Monte Carlo results (Fig.~1d).}
\vskip 12pt\noindent
{\bf Fig.~2} {$Z'$ Reservation for model $A$, for $r=-10$.}
\vskip 12pt\noindent
{\bf Fig.~3} {The same as Fig.~2 for $r=10$.}
\vskip 12pt\noindent
{\bf Fig.~4} {The same as Fig.~2 for $r=-1$.}
\vskip 12pt\noindent
{\bf Fig.~5} {The same as Fig.~2 for $r=1$.}
\vskip 12pt\noindent
{\bf Fig.~6} {The same as Fig.~2 for $r=0$.}
\vskip 12pt\noindent
{\bf Fig.~7} {$Z'$ Reservation for AGC model (model $B$).}
\vskip 12pt\noindent
{\bf Fig.~8} {Region for the $Z'$ lepton couplings ($g'_{Vl}, g'_{Al}$)
that would produce a visible $2 \sigma$ effect on $\sigma_b$ (Fig.~8a)
and on $\sigma_5$ (Fig.~8b). The two solid curves correspond to 
the two representative couples 
of $\xi$ quark couplings $\xi_{Vb} = \xi_{Ab} = 2$ and $\xi_{Vb} 
= \xi_{Ab} = -2$ and are compared with the ``leptonic channel 
ellipse'' (dashed curve). QED corrections are included. }

\newpage
\vspace*{-3.5cm}
\hspace*{-3cm}
\epsfig{file=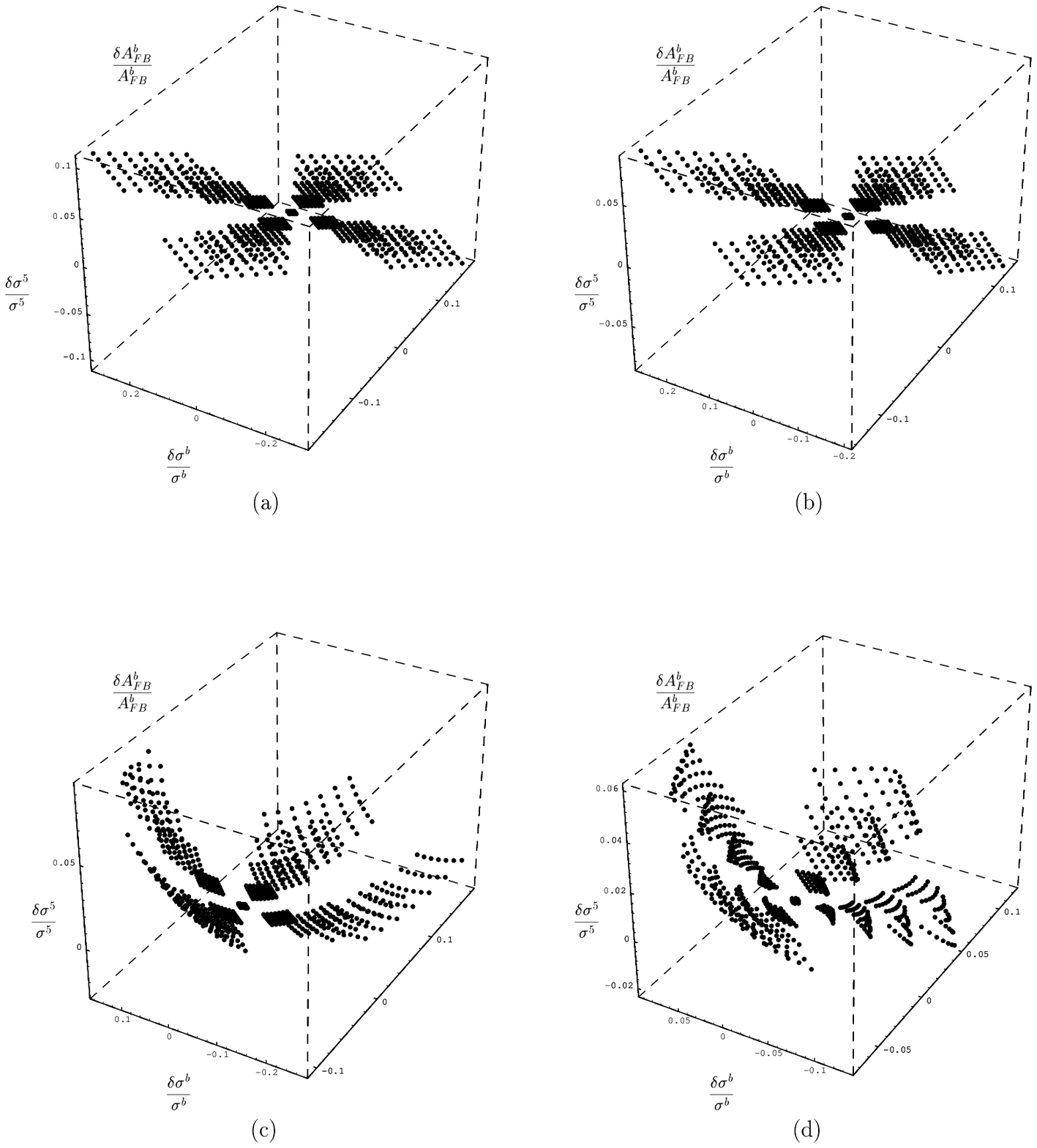}

\vspace*{-5cm}
\centerline{ {\Large Fig 1 }}

\newpage
\vspace*{-3.5cm}
\hspace*{-3cm}
\epsfig{file=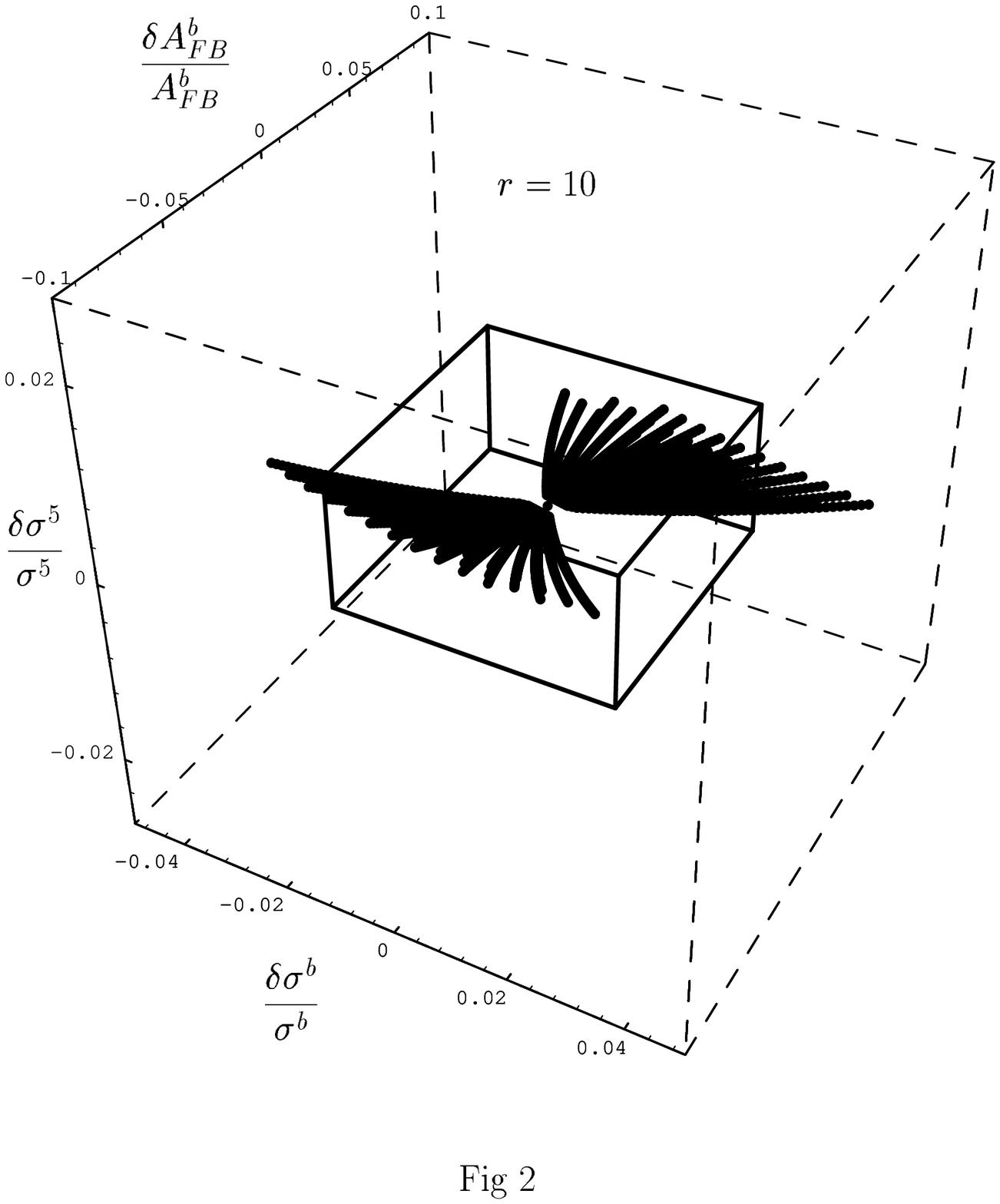}

\newpage
\vspace*{-3.5cm}
\hspace*{-3cm}
\epsfig{file=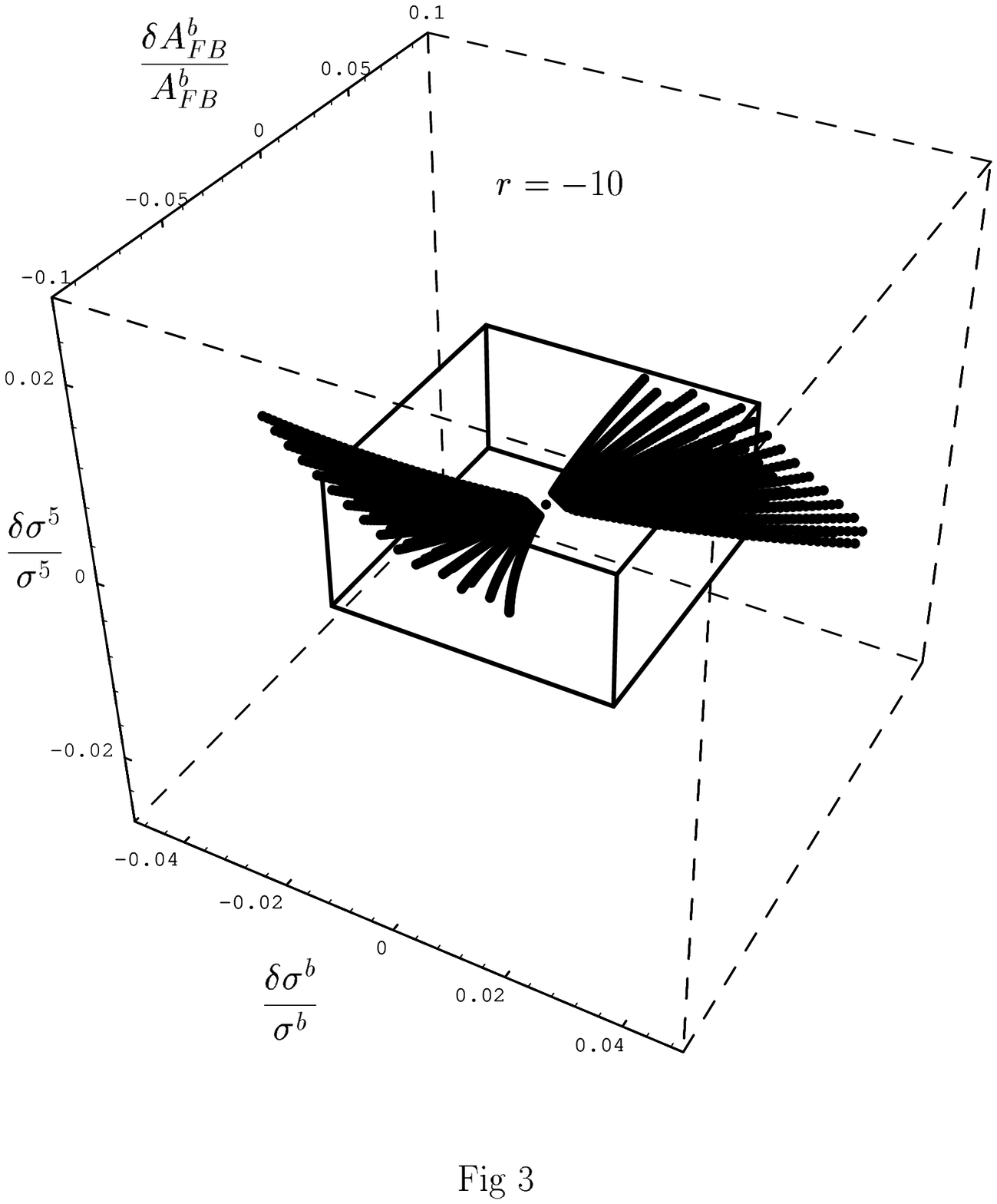}

\newpage
\vspace*{-3.5cm}
\hspace*{-3cm}
\epsfig{file=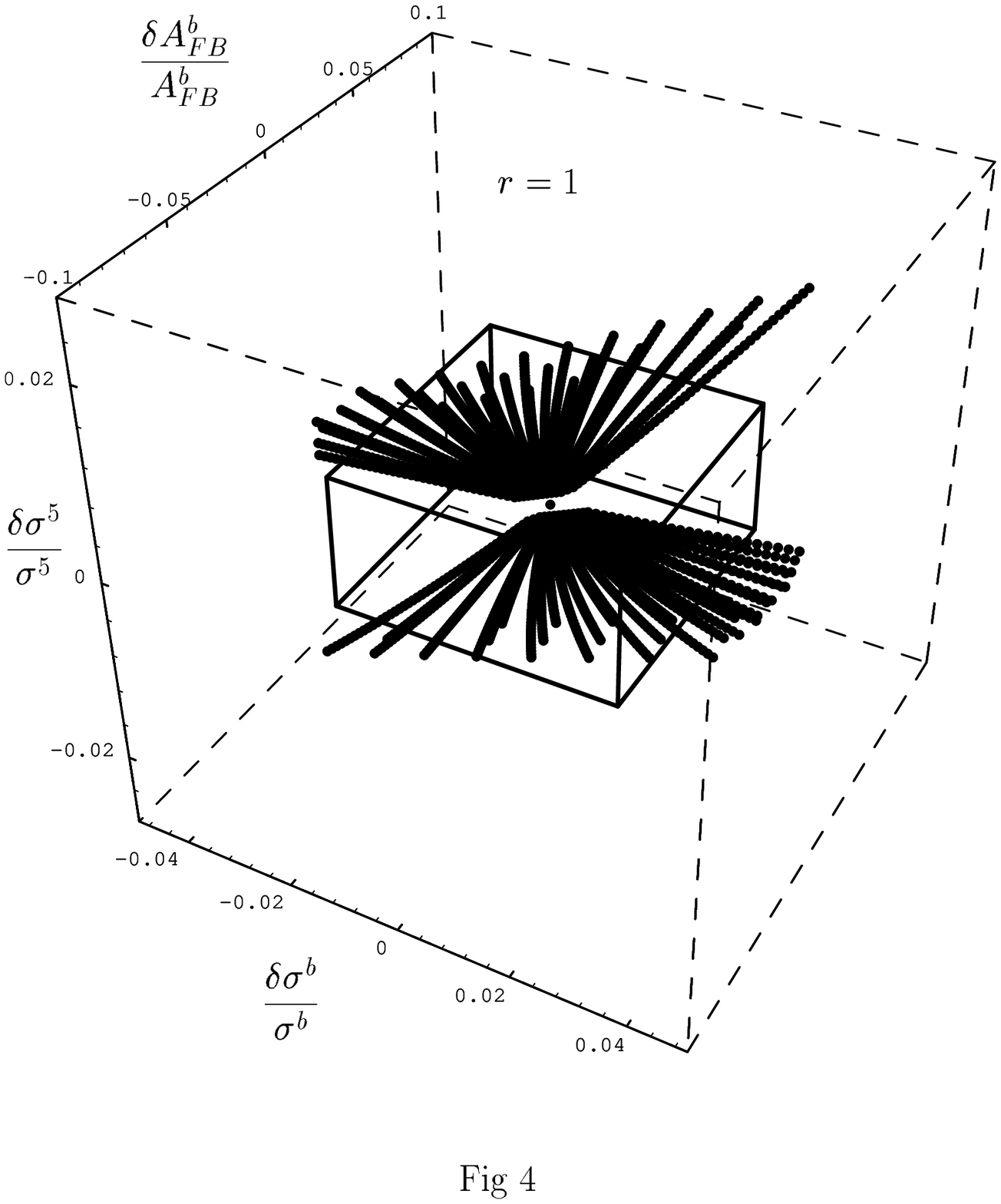}

\newpage
\vspace*{-3.5cm}
\hspace*{-3cm}
\epsfig{file=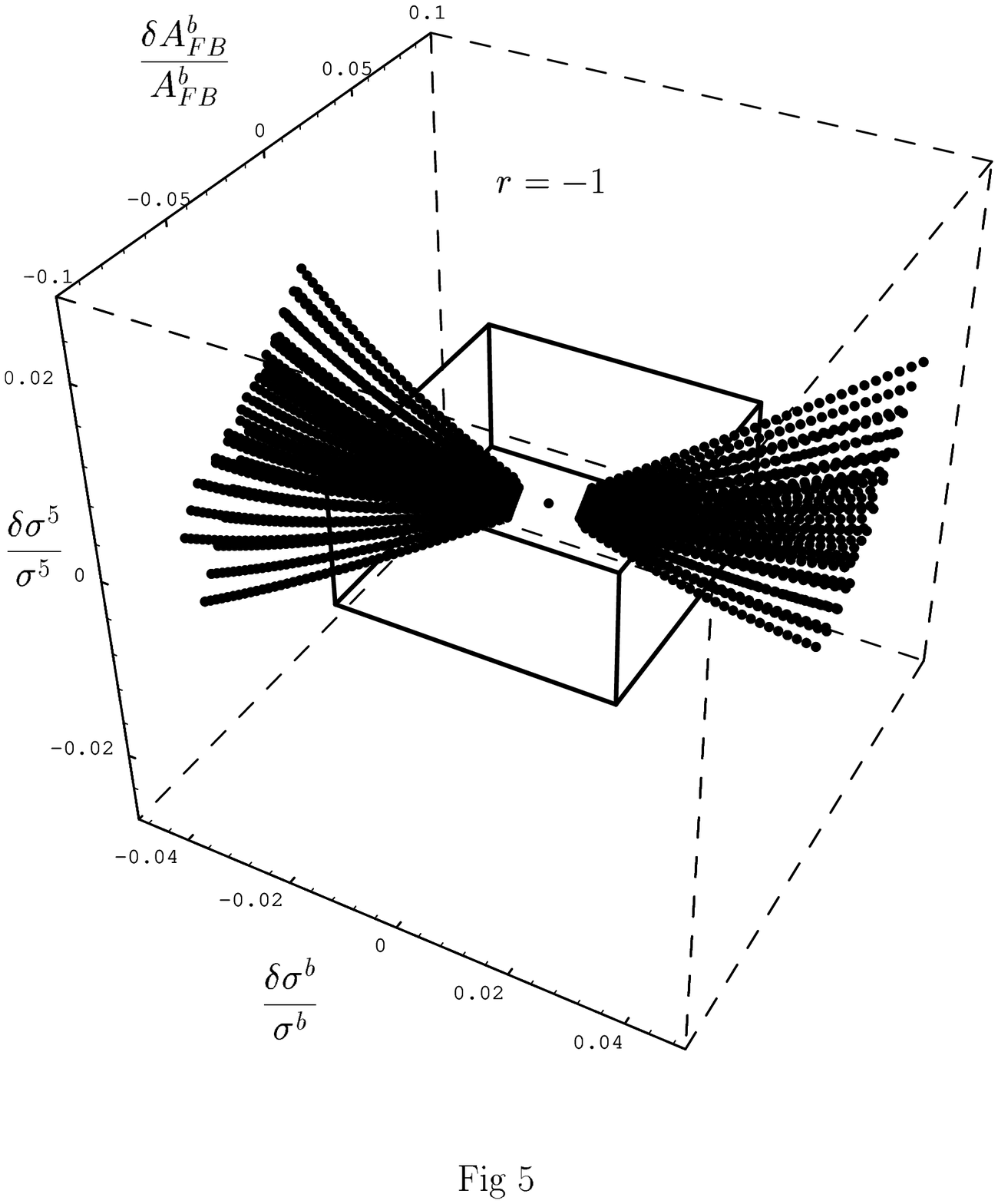}

\newpage
\vspace*{-3.5cm}
\hspace*{-3cm}
\epsfig{file=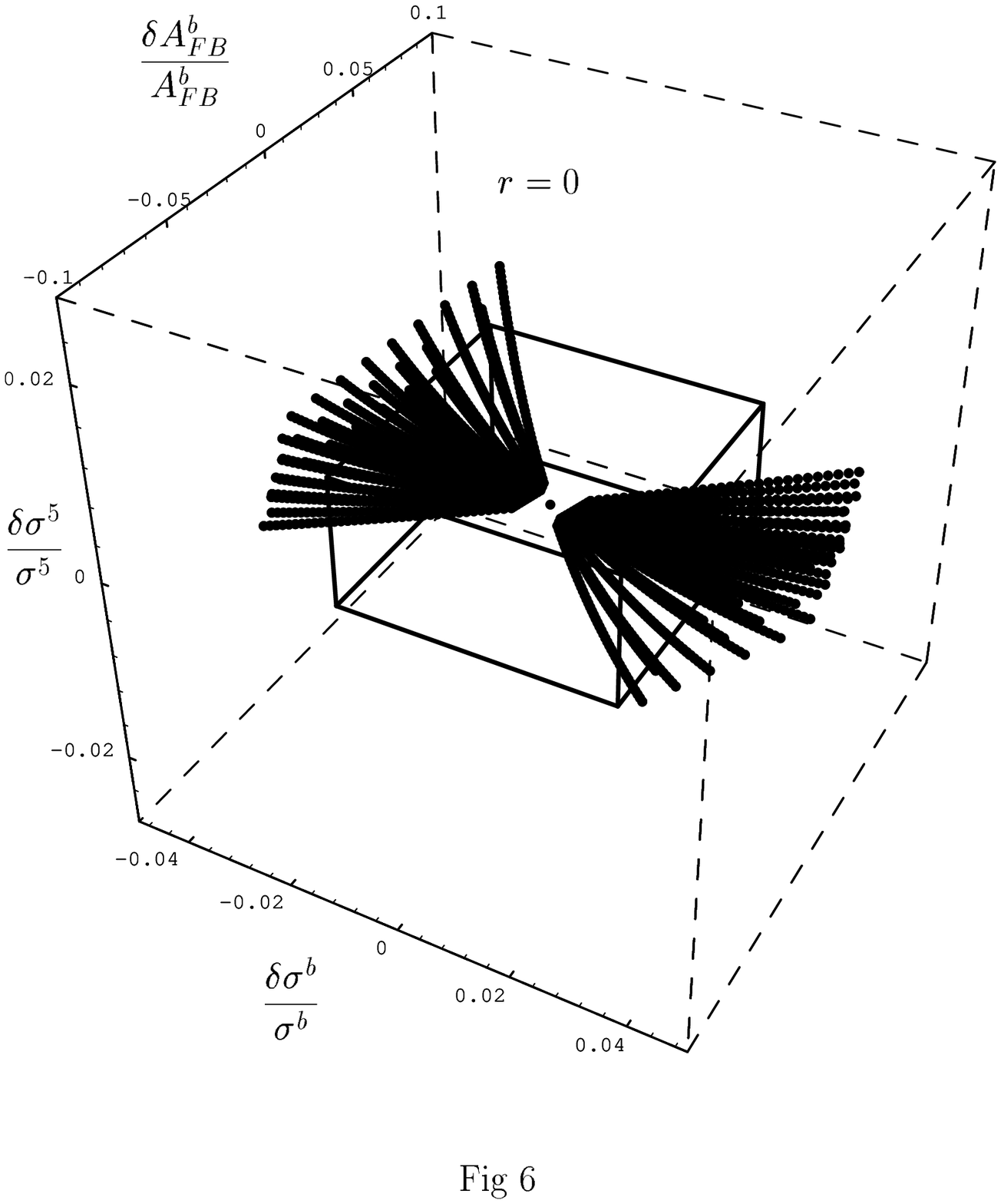}

\newpage
\vspace*{-3.5cm}
\hspace*{-3cm}
\epsfig{file=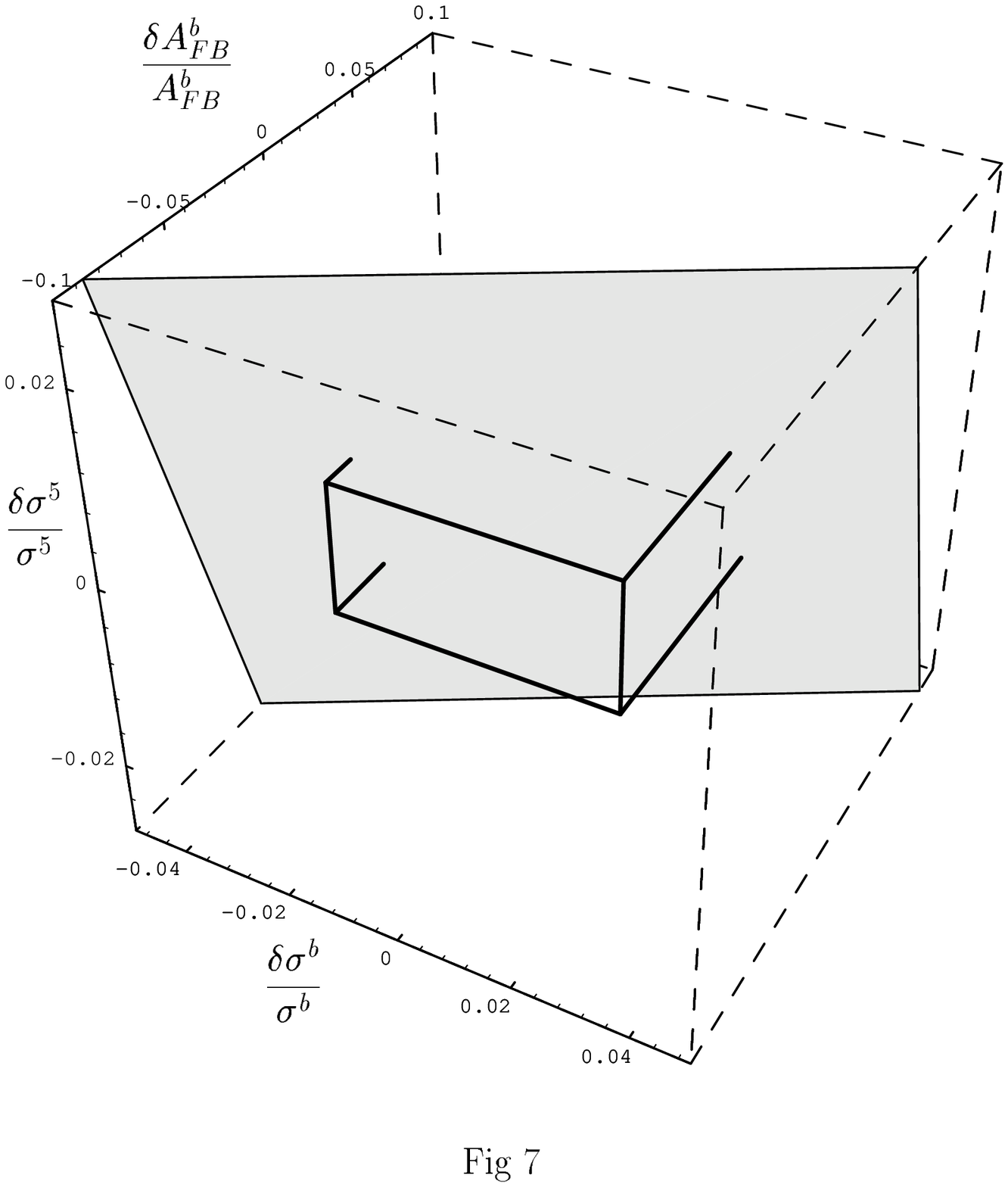}

\newpage
\vspace*{-3.5cm}
\centerline{ \epsfig{file=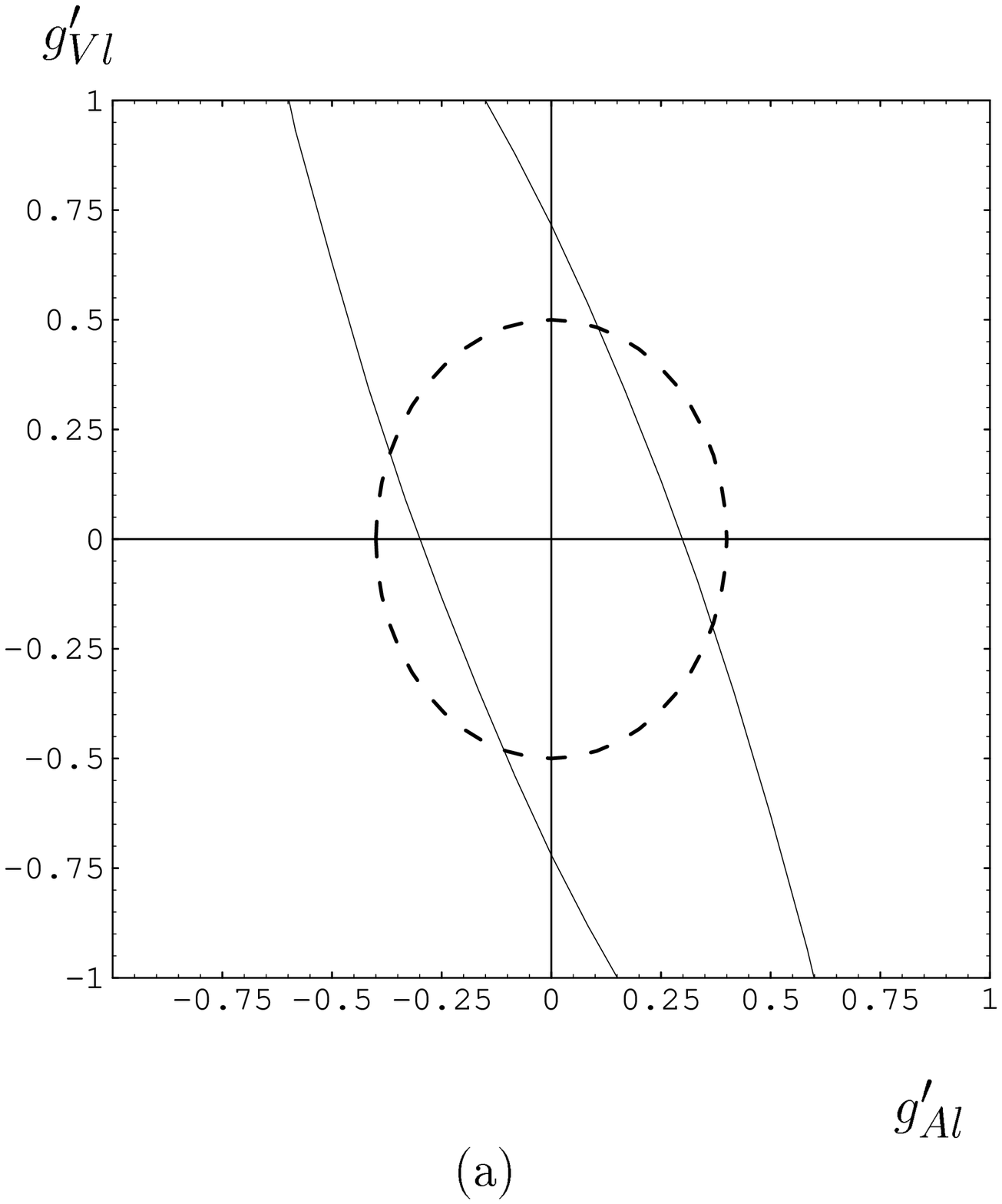,height=14cm} }
\vspace*{-2.5cm}
\centerline{\epsfig{file=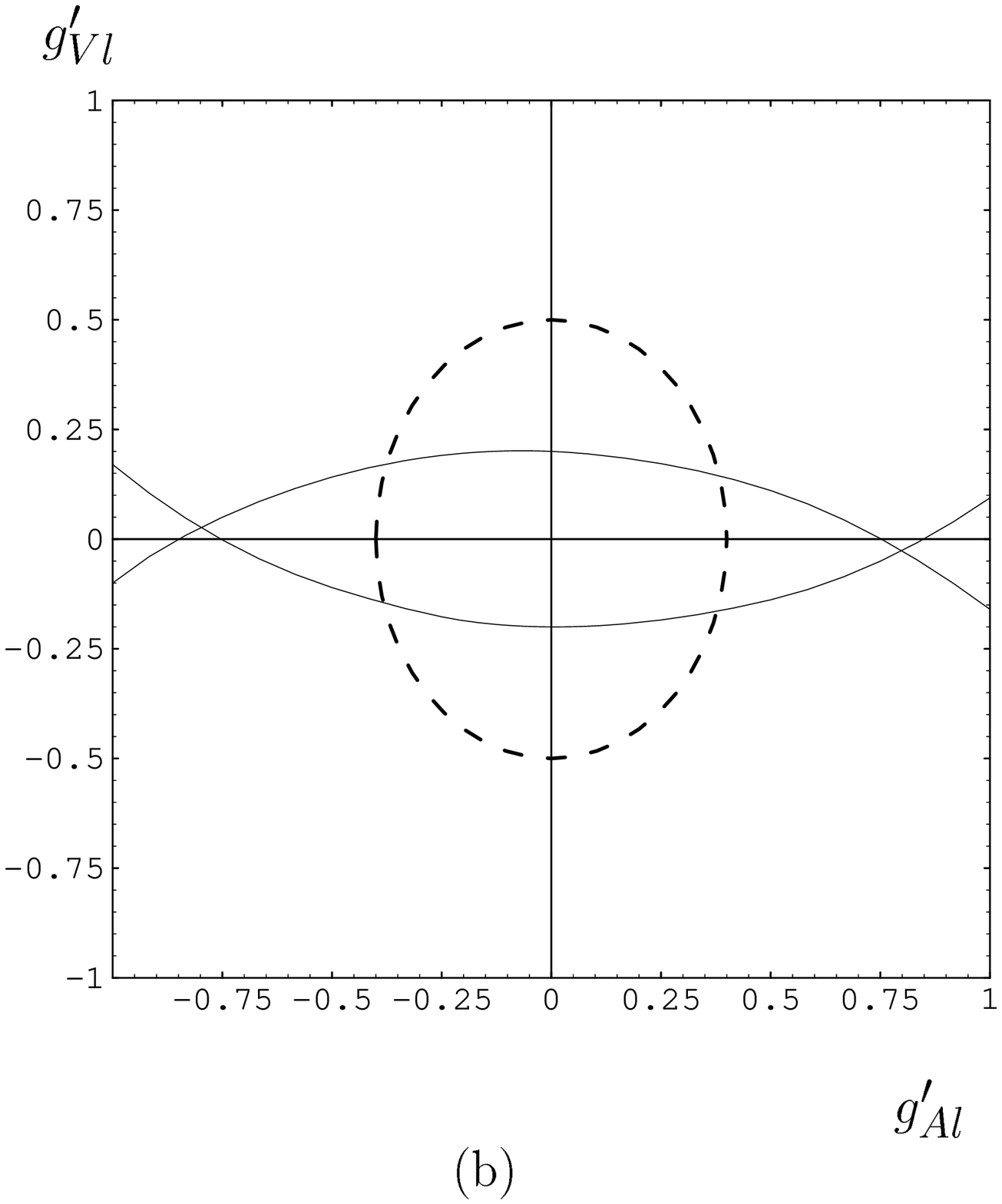,height=14cm} }
\vspace*{-2.5cm}
\Large \centerline{Fig 8 }

\end{document}